# Femtosecond Photophysics of Molecular Polaritons

*Francesca Fassioli,*[†,‡,⊥] *Kyu Hyung Park,*[†,⊥] *Sarah E. Bard,*[†] *Gregory D. Scholes*[†,*]

[†]Department of Chemistry, Princeton University, Princeton, New Jersey 08544, United States

[‡]SISSA – Scuola Internazionale Superiore di Studi Avanzati, Trieste 34136, Italy

[⊥]Contributed equally to this work

AUTHOR INFORMATION

**Corresponding Author**

*email: gscholes@princeton.edu

**ABSTRACT** Molecular polaritons are hybrid states of photonic and molecular character that form when molecules strongly interact with light. Strong coupling tunes energy levels and importantly, can modify molecular properties (e.g. photoreaction rates) opening an avenue for novel polariton chemistry. In this perspective, we focus on the collective aspects of strongly coupled molecular systems and how this pertains to the dynamical response of such systems, which though of key importance for attaining modified function under polariton formation, is still not well understood. We discuss how the ultrafast time and spectral resolution make pump-probe spectroscopy an ideal tool to reveal the energy transfer pathways from polariton states to other molecular states of functional interest. Finally, we illustrate how analyzing the free (rather than electronic) energy structure in molecular polariton systems may provide new clues into how energy flows and thus how strong coupling may be exploited.

*Introduction.* When an optical transition in matter couples to light in the strong coupling regime, new hybrid states - a quantum superposition of matter and photon degrees of freedom - are formed. These so-called *polaritons* give rise to new energy levels and therefore have the potential to modify the properties of the strongly coupled matter. Strong light-matter coupling has been extensively studied in the context of Rydberg atoms[1,2] and inorganic semiconductors since its first demonstration in 1975[3]. Strong coupling between organic molecules and light in optical cavities emerged years later, stimulated partly by the experimental work by Lidzey et al.[4] and Fujita et al. in 1998.[5] However, it has been only recently that the study of *molecular polaritons* has gained momentum thanks to the work by the Ebbesen group that has put forward the idea that chemical and physical processes in organic materials may be modified and controlled by strong light-matter coupling.[6-14]

To realize strong coupling between light and molecules in experimental setups, molecules are usually placed inside a resonant optical cavity (e.g. Fabry-Perot cavity) (Figure 1a) or coupled to a resonant surface plasmon.[15-17] Strong light-matter coupling is achieved when the coupling between the molecules and the optical mode divided by $\hbar$ is faster than the intrinsic decay of molecular excited states and photons. Hybridization of molecules with photons can readily be inferred by looking at how the molecular absorption changes inside an optical cavity (Figure 1b).

Typically, absorption at the resonant molecular transition disappears and splits into two peaks located below and above the absorption peak of bare molecules. These peaks arise from the formation of the so-called *lower polariton* (LP) and *upper polariton* (UP) transitions.

The energy difference between the LP and UP states defines the *Rabi splitting* $\Omega_R$ (Figure 1c). The magnitude of the Rabi splitting can reach values of orders up to 1 eV for electronic strong coupling,[18] and be even comparable to the molecular transition energy,[13,18-20] reaching what is known as the *ultrastrong coupling* regime (defined as a Rabi splitting above 10% of the transition energy).[21-24] Such large Rabi splittings are remarkable considering that the typical interaction strength of a single molecule to an optical mode is much smaller, on the orders of 0.01 meV ($10^{-2}$ cm$^{-1}$) for IR vibrations[25] and 1 meV (1 cm$^{-1}$) for electronic transitions in the visible spectrum.[26]

The single molecule to light coupling is also much smaller than typical molecular line broadening, which means that single molecules usually couple weakly to light. Yet, strong light-matter coupling can easily be achieved at ambient conditions in organic molecules. What allows strong light-matter coupling in molecular systems is the possibility of using a high concentration of molecules that couple to the same effective optical mode, thereby producing a collective state. Transitions from the ground state to this collective state (e.g. the LP state) have greatly increased dipole strength compared to the transitions of each molecule individually. In the LP and UP states, the molecules no longer absorb and emit light independently—as they would in free space in the absence of intermolecular interactions—but instead do so collectively owing to their indirect interaction via a common mode of the electromagnetic field. The regime of strong coupling can then be achieved even in the absence of strong coupling for a single molecule, relaxing the need of using cavities with high finesse (very low cavity photon loss) or low temperature conditions, which is normally a requirement in inorganic and atomic microcavities.[27-30]

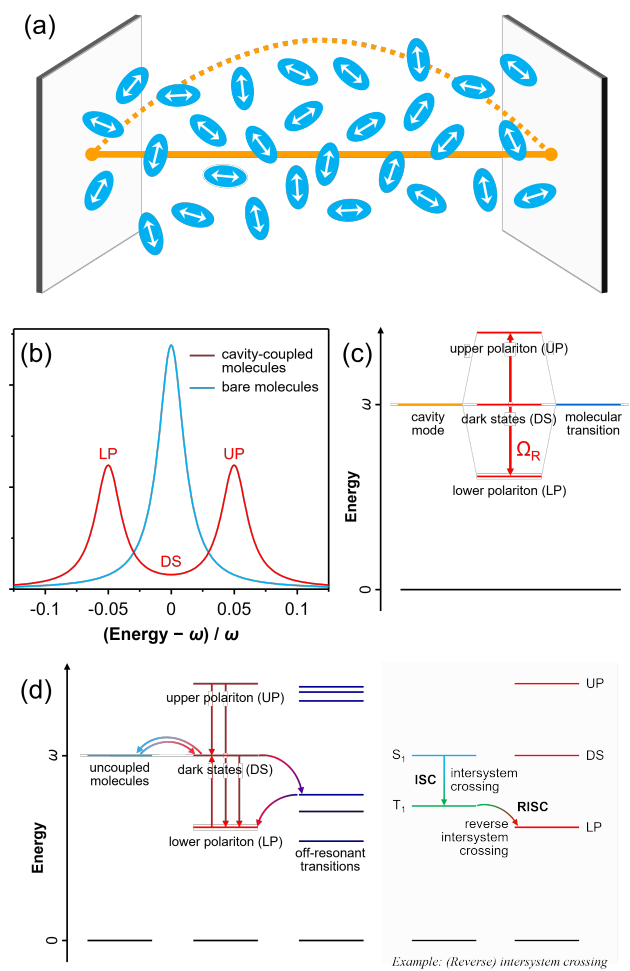

**Figure 1.** (a) Schematic representation of molecules strongly coupled to an optical mode of a microcavity. Transition dipole moments of the molecules are shown as arrows. (b) Absorption spectra of bare (cyan) and cavity-coupled (red) molecules. (c) Energy diagram of cavity mode (yellow), molecular transition (blue), and polariton states (red), where energy difference between upper (UP) and lower (LP) polariton states is defined as Rabi splitting, $\Omega_R$. (d) Interaction between polariton states (red) and reservoir states (cyan) or other uncoupled states (dark blue). The energy transfer between these states is indicated as arrows. In the case of interaction between polariton states and, for example, triplet states, the dynamics manifest as modified (reverse) intersystem crossing.

The collective strong coupling of many molecules to light implies that not only the energy landscape of the molecular system is modified, but also importantly, that macroscopic polariton states, which correlate large numbers of distant molecules that otherwise would not interact, may be formed. This offers a resource to be exploited for modifying molecular function through strong coupling. Several studies have already demonstrated that it is indeed possible to change the properties of chemical systems by polariton formation. For example, electronic strong coupling has been proposed to modify exciton transport and enhance conductivity in an organic semiconductor,[8,10,13,31] while vibrational strong coupling in solution phase apparently modifies the rate of chemical reactions by changing the energetics of the reaction coordinate in the electronic ground state.[9,12,32,33] Although these and other studies have provided intriguing evidence that the properties of a material can be modified by strong coupling, many fundamental aspects of the photophysics of molecular polaritons are yet to be understood. In particular, while the energy levels of the emerging light-matter system can be predicted from theoretical models with relative success, how polaritons interact among themselves and with other states is much less clear, making it difficult to predict how polariton formation might ultimately be used to control excited state processes.

For example, an open question is how polaritons will interact with purely molecular states (Figure 1d). Among these are molecular states at the same energy of the resonant transition that is being coupled to the optical mode, which may have two origins. The first are states of molecules that do not couple to the optical mode forming what is commonly referred to as a *reservoir* of uncoupled molecules, and arise under conditions where only a small fraction of molecules couples to light and hence participates in polariton formation.[34-38] The second kind corresponds to purely molecular states that arise as eigenstates of the combined light-matter system, even when all molecules in the sample couple to the optical mode of the cavity.

When many molecules are involved in polariton formation, the symmetry of the interaction with the optical mode results not only in the polariton (hybrid light-matter) states but also in a large number of purely molecular *dark states* located at the energy of the bare molecules.[39-42] Due to their high-density, dark states can act as a sink of excitation from polaritons, and similarly, polaritons may transfer energy to the uncoupled molecules.[36,37,40,43] It is therefore important to understand the conditions under which the response of a system is defined by polariton dynamics rather than dominated by uncoupled molecules or dark states.

Molecular transitions that are not resonant with the cavity may also be involved in the response of a system. It is often the case that the molecular property that one aims to control involves such states that do not directly couple to the optical mode. These off-resonant states are qualitatively different from the reservoir states, which are resonant yet do not couple due to the improper orientation of their transition dipole moment or positioning. One such example includes the works aimed at modifying the reverse intersystem crossing rate (RISC) between a triplet and singlet state by coupling the singlet state to the cavity mode (Figure 1d right).[26,44-46]

In this perspective, we will discuss the collective aspects of molecular polaritons followed by how polaritons interact with other energy states in the system (e.g dark states). In order to understand the dynamical processes taking place in strongly coupled molecular systems and how polaritons interact with other purely molecular states, we need to be able to 'follow' polaritons in an ultrafast timescale after photoexcitation. For this reason, a few groups including ours are resorting to femtosecond pump-probe spectroscopy to study molecular polaritons. With the aid of some recent works by our own and other groups, we will illustrate how spectroscopies, in particular pump-probe spectroscopy, can give unambiguous evidence of polariton formation and unveil the dynamical evolution and energy transfer pathways in organic microcavities.

***Synchronizing distant molecules: the energy structure in the Tavis-Cummings Hamiltonian.*** In order to describe the emergence of hybrid light and matter energy states and understand how strong light-matter coupling may modify chemical processes, a full quantum treatment of the properties of light and matter is necessary. We note that while classical or semi-classical approaches (e.g. the classical transfer matrix method and semi-classical F-matrix fitting method respectively) suffice to predict linear optical spectra of strongly coupled systems, a fully quantum treatment is needed to describe observables such as cavity photon statistics and most importantly, the interaction between polariton states and other molecular states.[47] We follow a 'few level' effective quantum Hamiltonian approach in which only the most relevant states are explicitly included. Here we introduce the simplest such model, namely the well-known Dicke model in quantum optics that describes *N* non-interacting two-level systems interacting with a single common optical mode.[48-50]

Even though this Dicke model has been traditionally used to describe atoms in optical cavities, each two-level system can be representative of the molecular dipole allowed transition that strongly couples to the cavity mode (Figure 1a). Albeit simple, it illustrates the main features of a strongly coupled system, that include the nature of the emergent (hybrid) energy states that are predicted from diagonalizing the light-matter Hamiltonian, and how to think about the photophysics and optical transitions in such a system. For a treatment of molecular polaritons that take into account the molecular vibrational structure, the reader is referred to the Holstein-Tavis-Cummings model.[51-57]

The Dicke Hamiltonian is given by

$$H = E \sum_{i=1}^{N} \sigma_i^+ \sigma_i^- + \omega a^\dagger a + g \sum_{i=1}^{N} (\sigma_i^+ + \sigma_i^-)(a + a^\dagger).$$

(Eq. 1)

$\sigma_i^+$ and $\sigma_i^-$ are molecular operators that create and destroy an electronic excitation on molecule $i$ respectively, $E$ is the energy of the dipole allowed transition, $a^\dagger$ ($a$) describe the creation (annihilation) operators of the bosonic optical field, $\omega$ is the frequency of the optical mode and $g$ measures the strength of the light-matter interaction. The interaction terms proportional to $\sigma_i^+ a$ ($\sigma_i^- a^\dagger$) describe molecular excitation(relaxation) by the absorption(emission) of a cavity photon, thus conserving the number of excitation quanta in the system. Instead, the terms proportional to $\sigma_i^+ a^\dagger$ ($\sigma_i^- a$) simultaneously create(destroy) a molecular excitation and a photon. The rotating wave approximation consists in neglecting these *counter-rotating* terms and is valid under resonance conditions and as long as the effective light-matter interaction, which scales with the number of molecules as $g\sqrt{N}$, is small in comparison to the molecular transition energy and frequency of the cavity mode, $g\sqrt{N} \ll E, \omega$. The resulting Hamiltonian, known as the Tavis-Cummings model, is given by

$$H_{RWA} = E \sum_{i=1}^{N} \sigma_i^+ \sigma_i^- + \omega a^\dagger a + g \sum_{i=1}^{N} (\sigma_i^+ a + \sigma_i^- a^\dagger).$$

(Eq. 2)

By neglecting the counter-rotating terms in the Hamiltonian in Equation 1, the total number of particles $N_p = \sum_{i=1}^{N} \sigma_i^+ \sigma_i^- + a^\dagger a$ (either molecular or photonic excitations) is conserved, i.e. $[H, N_p] = 0$. As a consequence, the light-matter Hamiltonian eigenstates comprise a superposition of states with the same number of excitations. From this it can be immediately inferred that the ground state is the state with all molecules in the ground state and no cavity photons which we denote by $|G, 0\rangle$. Eigenstates with $N_p > 0$ define a ladder of states bound by a lower polariton LP($N_p$) and upper polariton UP($N_p$) separated in energy by the Rabi splitting $\Omega_R(N_p)$.[27,58,59]

Under low intensity photoexcitation, which is often used in experiments, only one- ($N_p= 1$) and two-particle ($N_p= 2$) eigenstates are accessed in pump-probe, the latter which can be populated through excited state absorption (ESA) from the one-particle states. Within the one-particle subspace, the light-matter interaction (Equation 2) induces transitions between the state with a single photon and no molecular excitations ($|G, 1\rangle$) and all states with a single molecular excitation (on site $i$) and no photons $|e_i, 0\rangle$. This gives rise to the following one-particle lower polariton (LP) and upper polariton (UP) states (Figure 1c) that take the form

$$|UP/LP\rangle = \frac{1}{\sqrt{2}}|G, 1\rangle \pm \frac{1}{\sqrt{2N}} \sum_{i=1}^{N} |e_i, 0\rangle,$$

(Eq. 3)

where we have assumed resonance conditions ($\omega = E$). The UP and LP have energies $E_{UP/LP} = \omega \pm g\sqrt{N}$ and define the *vacuum Rabi splitting* as $\Omega_R = E_{UP} - E_{LP} = 2g\sqrt{N}$. These are the eigenstates that are visible in low-intensity absorption and transmission/reflection experiments as two peaks at the energies of the LP and UP and are separated by $\Omega_R$ (Figure 1b,c). It is remarkable that even in the absence of intermolecular coupling polaritons delocalize over all molecules due to the interaction to the common optical mode, mediating an indirect molecule-to-molecule interaction.

In addition to the two polariton states, the light-matter coupling also gives rise to a set of *N*–1 degenerate dark states located at the energy of the isolated molecules, $E = \omega$ (Figure 1c). Given that in most realizations, a large number of molecules is required to achieve strong light-matter coupling, a large density of dark states is predicted. The dark states (DS) are purely molecular exciton states of the form $|DS\rangle = \sum_{i=1}^{N} c_i |e_i, 0\rangle$, with coefficients satisfying $\sum_{i=1}^{N} c_i = 0$. As their name suggests, dark states carry a vanishing dipole moment from the ground state. Dark states are slightly delocalized in such a way that phases of wavefunctions in the molecule basis cancel, thereby ensuring that transitions from the ground state to the DS are forbidden. This has to be the case because photonic modes at the mirror boundaries are forbidden at the cavity resonance wavevectors.

*Coherence, disorder.* The Tavis-Cummings Hamiltonian clearly illustrates the remarkable delocalization of polariton states. However, from the theory of molecular excitons it is well known that energetic disorder can cause excited states to localize, so it is of interest to determine how robust polariton states are with respect to disorder in the transition energies of the molecules. In particular, since line broadening of molecular transitions is strong, we should work out how much mixing this disorder introduces to the polariton states.

In recent work,[60] the interplay of disorder and light-matter coupling was explored. We found that disorder broadens the high density of dark states centered at the energy of the bare molecules (Figure 2a–2c). Even under relatively high energetic disorder compared to the light-matter coupling strength in a resonant cavity, polariton bands are strongly delocalized. The robustness of the lower polariton state delocalization is because the LP state is strongly split away from the other states, which impedes the usual mechanisms for decoherence and localization.

The incredibly long-range and robust delocalization is particular to the Hamiltonian for the star graph[60] (i.e. the Tavis-Cummings model confined to the single-excitation subspace). Moreover, the polariton bands are not broadened commensurate to the disorder (e.g. compared to the dark state density of states). This bandshape narrowing is attributed to exchange narrowing, which arises because the coherent exciton delocalization means the system averages over many configurations of energy disorder—the slow random fluctuations at each molecular site—on a timescale faster than the measurement of the lineshape. This averaging narrows the absorption lineshape by canceling oppositely signed fluctuations[61].

In other work,[62] we have studied the delocalization of polariton states in various ways. To examine delocalization we need to ascertain how different the state of interest is from the corresponding pure state. In other words, the measures indicate how 'mixed' the state is owing to averaging over the ensemble of configurations of the system. We chose to assess the mixedness using three measures (Figure 2d). These are (i) the inverse of the inverse participation ratio 1/IPR, often used as a measure of exciton delocalization; (ii) the purity, which is 1 for a pure state and diminishes as a state becomes more mixed; (iii) the relative entropy of coherence, which measures how far (entropically) the state is from the nearest fully mixed state.

The analyses of mixedness were carried out by analyzing the ensemble of eigenfunctions obtained from the Tavis-Cummings Hamiltonian, with disorder in the diagonal elements that represent molecular transition frequencies.[60] To quantify the mixedness of the delocalized molecular component of the wavefunctions only, the photonic contribution to the wavefunctions was traced out. The $N$ molecule-only eigenstates, then, are indexed by $j$, with a basis of single-molecule excitation states (denoted by the electronic excited state at site $n$):

$$\Psi^j = \sum_{n=1}^{N} a_n^j |n\rangle \tag{Eq. 4}$$

Ensemble properties are estimated using Monte-Carlo sampling. Each independent realization of an exciton is constructed by adding energy offsets, from a Gaussian random distribution, to the diagonal elements of the Hamiltonian matrix[63].

The delocalization of excitation in an ensemble of multichromophore arrays can be quantified by the Inverse Participation Ratio (IPR)[64], which is an effective measure for simulations where exciton wavefunctions are ensemble-averaged[65-67]. IPR characterizes the variance of probabilities within a wavefunction delocalized among $N$ molecule-local sites. The IPR for exciton state $j$, $I_j$, is defined by Equation 5 and we plot in Figure 2 the inverse of the IPR as a function of cavity coupling strength, which indicates the delocalization length in number of molecules.

$$I_j = \sum_{n=1}^{N} \left| a_n^j \right|^4 \tag{Eq. 5}$$

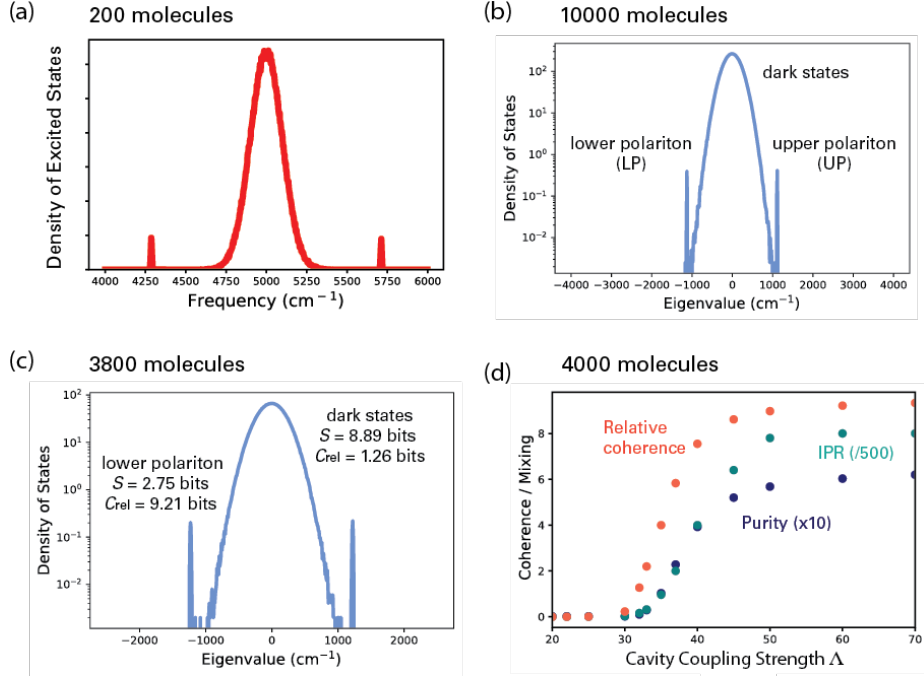

**Figure 2.** (a) Spectrum of the ensemble single-excitation density-of-states calculated for 200 molecules coupled to a resonant cavity mode, with coupling strength $g = 50$ cm$^{-1}$. The transition energies of the molecules are randomly disordered, standard deviation 100 cm$^{-1}$. (b) A simulation for 10000 molecules, 500 samples of the ensemble, energy disorder in the molecular lineshape 200 cm$^{-1}$, and (unitless) cavity strength factor $\Lambda = 40$, which corresponds to a peak coupling strength $g = 18$ cm$^{-1}$. The vector potential inside a cavity scales with the cavity strength factor, and therefore the light-matter coupling strength is proportional to the cavity strength factor (see equations 2 and 3 in ref. 62). (c) Simulation as in part b, but with 4000 molecules in the ensemble, 3800 of which couple to the effective cavity mode of the Tavis-Cummings model. The cavity strength factor $\Lambda = 60$. (d) Relative coherence measure $C_{rel}(\rho)$, inverse of the inverse participation ratio (IPR), and purity calculated for the lowest eigenstate ensemble (the LP state) as a function of cavity strength factor $\Lambda$ for ensembles of 4000 molecules. The simulations include energy disorder in the molecular lineshape with standard deviation 200 cm$^{-1}$. Calculations of plots b-d are described in ref 62.

Ensemble density matrix operators are defined in the molecular site basis according to,

$$\rho_{mn}^{j} = \left\langle a_m^j a_n^j \right\rangle_M \quad \text{(Eq. 6)}$$

where the angle brackets indicate that an average of the density matrix is taken over $M$ realizations of the ensemble, and $j$ denotes the index of the eigenstate(s) being considered. The purity is defined as Tr $\rho^2$. It is 1 for a pure state, and less than 1 (but positive) for a mixed state.

The loss of exciton delocalization of eigenstate $j$, that is, its degree of mixed character in the ensemble average[68-72], is estimated from the density matrix using the von Neumann entropy $S_j$, in units of bits:

$$S_j = -\text{Tr}\,(\rho \log_2 \rho) \qquad (\text{Eq. 7})$$

The von Neumann entropy has a value of 0 for a pure state, and $\log_2 N$ for an ensemble density matrix that is fully mixed. The von Neumann entropies of the LP band and the peak of the dark state band are compared in Figure 2c. The von Neumann entropy is low for the LP state because the state is not very mixed in character. Indeed, the closest fully mixed state is 9.2 bits distant from this LP band. The dark state band is not fully mixed, but is 1.3 bits distant from the closest fully mixed state.

We have used a distance measure of coherence, the relative entropy of coherence, to assess how far states are from the nearest fully mixed state[73]. The relative entropy of coherence $C_{\text{rel}}$ is defined by Plenio and co-workers as the difference between the von Neumann entropy of the state, with density matrix $\rho$, and that of the density matrix $\sigma$, where the off-diagonal elements of $\rho$ have been set to zero. Here, $\sigma$ is the closest fully mixed density matrix.

In Figure 2d, the mixedness measures of inverse of the IPR, purity, and relative entropy of coherence are plotted as a function of the coupling strength of the cavity, for an ensemble of 4000 molecules for the lowest eigenstate in the ensemble (LP). The cavity coupling strength indicates the cavity Q-factor, and it affects the sharpness of the vector potential mode function in the cavity as well as the peak radiation-matter coupling. The collective coupling 'turns on' abruptly for a critical value of the cavity coupling strength, depending on the number of molecules in the cavity.[62] This 'turn on' of the collective coupling is reflected by the mixedness measures.

***Modifying excited state dynamics by strong coupling.*** It is clear that strong-light matter coupling can give rise to new states of the coupled system that differ both in energy and nature from those of the uncoupled system. But how is the dynamical response of the system modified by these new states? For example, how does the polariton lifetime compare to that of the molecular excited state outside of the cavity? Since polaritons have both a light and matter component, one could expect their lifetime to be limited by the shortest lifetime between the molecular excited state and the cavity photon. The lifetime of molecular excitons varies widely from hundreds of femtoseconds to milliseconds but is generally orders of magnitude longer than the typical cavity photon lifetime, which is in the order of tens of femtoseconds (possible exceptions include the coupling to a higher energy state with a shorter lifetime). Several studies have indeed reported polariton lifetimes matching the order of magnitude of the cavity photon lifetime.[38,74,75] Interestingly, other works have instead shown that polaritons can survive much longer than the cavity photon lifetime and even longer than the lifetime of the molecular exciton.[26,36,43,76] Such long polariton lifetimes can be explained as arising from either an intrinsic long lifetime of the polariton state or from energy transfer between polaritons and other molecular excited states (e.g. dark state manifold or non-resonant higher or lower excited states of the molecules).[36,76-79]

Intrinsic polariton lifetimes that do not simply relate to their matter and photon constituents can be expected when the light-matter coupling is strong enough. In this case, the system relaxes in the dressed basis, and decay and decoherence rates are weighted by the spectral density of the electromagnetic/molecular environment evaluated at the new energy gaps of the hybrid states.[40,80,81] Polariton lifetimes that are instead limited by energy transfer with other molecular states have also been proposed, for example by Lidzey and coworkers, who have shown that polaritons may exchange energy with the vibrationally excited state of uncoupled molecules, which could explain for instance the observation of anti-Stokes fluorescence after photoexcitation of the lower polariton.[37,82] Distinguishing between these two possible mechanisms that lead to long-lived polariton features is only possible with an experimental technique with an ultrafast fs time as well as frequency resolution that can unveil energy transfer pathways. In the next section, we will discuss how pump-probe spectroscopy can shed light into this and other open questions regarding the formation of polariton states and their dynamics in organic microcavities.

***Pump-probe spectroscopies can reveal energy transfer pathways and timescales in polariton systems.*** Transmission/reflection linear spectra can provide a first indication of polariton formation with the emergence of the characteristic splitting in the transmission peak of the optical cavity when a resonant material is placed inside it. While the magnitude of this splitting compared to the spectral widths of the empty cavity transmission and bare molecular absorption provides a strong indication of whether the strong light-matter coupling regime has been achieved, the same spectral shape can be expected from a purely classical phenomenon with no requirement of polariton formation.[19,83-85] It is therefore often necessary to also analyze the absorption spectra of the organic microcavity for stronger proof that new hybrid states have been formed.

Non-linear ultrafast spectroscopy offers several advantages over linear spectroscopy for probing organic microcavities (Figure 3). First, it provides dynamical information about the coupled system, such as revealing the emergence of new timescales that are absent in free-space, and second, it is able to probe the ladder of polariton states with different number of particles.[27,58,59] In

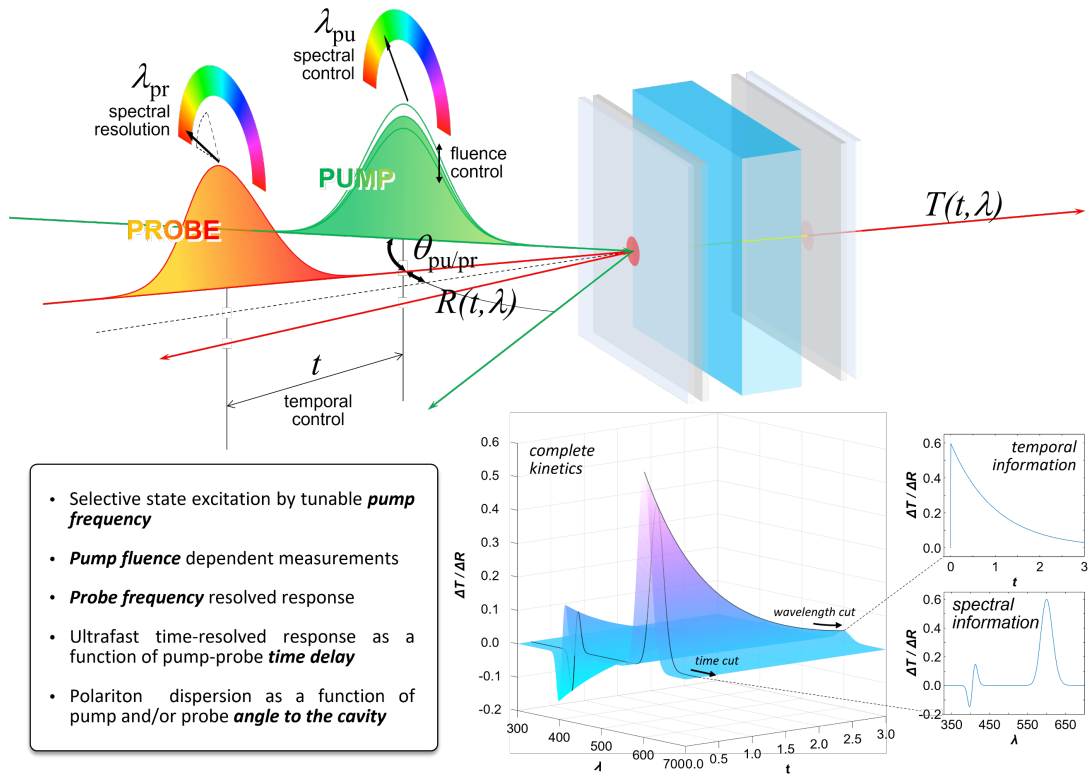

**Figure 3.** Schematic illustration of pump-probe spectroscopy on a strongly coupled cavity. Pump-probe spectroscopy has the advantage of controlling pump energy and fluence to track excited-state dynamics following selective excitation of a polariton state. Spectral and temporal information can be retrieved using a broadband probe, which allows us to study the ladder of polariton states and their interaction with uncoupled molecular states. this way, pump-probe spectroscopy can give unequivocal evidence of polariton formation as well as information on how the response of a system, and hence its physical and chemical properties, are modified by strong coupling.

In a pump-probe experiment, an ultrafast laser field – the *pump* – is first used to photoexcite the system, and a second pulse – the *probe* – is used to retrieve information about the state of the system after the action of the pump. In cavity coupled materials, pump-probe experiments measure either reflectance and/or transmittance after the action of a single pump pulse which is probed as a function of the delay time between pump and probe. Changes in reflectance and/or transmittance are detected in the *transient spectra,* which is obtained as the difference between the spectra after and before the pump pulse has acted upon the system (Figure 3, bottom). Ground state bleach (GSB) appears then as a negative feature in a transient spectra, while excited state absorption (ESA) as a positive one.

One of the advantages of this technique is that it is possible to *selectively* photoexcite a given excited state (e.g. LP or UP) by using spectrally narrow pump pulses. The response of the system can then be measured on an ultrafast timescale at different pump-probe time delays as a function of probe frequency. In this way, different energy transfer pathways can in principle be disentangled, providing precise temporal and frequency resolved information on the relaxation processes taking place after photoexcitation.

One of the first studies using transient ultrafast spectroscopy on organic microcavities came from the Ebbesen group, who studied the energy relaxation pathways in a strongly coupled J-aggregate.[36] By performing selective excitation of the UP, LP and uncoupled exciton states and then analyzing the transient spectra as a function of probe wavelength, they could unambiguously determine that energy exchange took place between polaritons and the exciton states of uncoupled molecules, and that importantly, the LP was intrinsically long-lived rather than its lifetime being limited by energy transfer to other states.

Despite the advantages of transient absorption measurements in organic microcavities, relatively few studies using this technique have been carried out so far, and therefore its full potential is yet to be appreciated. The aim of the next two sections is to illustrate more in detail the information that can be gained as well as the limitations of pump probe in studying organic microcavities. We will discuss two representative works, starting with the study of a molecular dye under electronic ultrastrong coupling that was carried out in our group, followed by the ultrafast transient response of vibration polaritons.

***Ultrafast dynamics of a molecular chromophore under electronic ultrastrong coupling.*** We studied the dynamical pump-probe response of charge-transfer dye molecules (4CzIPN) in an optical cavity, where the 460 nm (~2.7 eV) molecular transition, which appears as a shoulder in the absorption spectra of bare molecules (Figure 4a), was strongly coupled to a resonant cavity mode.[20] The high oscillator strength of the transition combined with a high concentration of molecules in the cavity, resulted in a very large Rabi splitting of 1.6 eV, about 60% of the molecular excitation energy, reaching the ultra-strong coupling regime (Figure 4b). This large Rabi splitting allowed for well-separated polariton peaks, free of spectral congestion, allowing to selectively photoexcite each polariton branch separately in pump-probe. The transient spectra obtained after resonant photoexcitation of the UP and LP (Figure 4c and 4d, respectively) were drastically different. Upon photoexcitation of the UP, a derivative-like shape appeared at the UP energy, while a positive feature appeared in the LP region. On the other hand, upon photoexcitation of the LP, we found a positive feature at the energy of the UP and a smaller positive feature at the energy of the LP. Such a large positive band in the UP region upon photoexcitation at the LP had not been reported previously.

Derivative-like features in cavity coupled molecular systems have previously been explained as arising due to a 'contraction' of the Rabi splitting after the pump, which reduces the density of molecules in their ground state and consequently the amount of molecules that are able to get excited and participate in polariton formation when the probe acts.[43,86-91] Such a situation though, appears under conditions in which the Rabi-splitting is dependent on the intensity of the pump.[88-90,92,93]

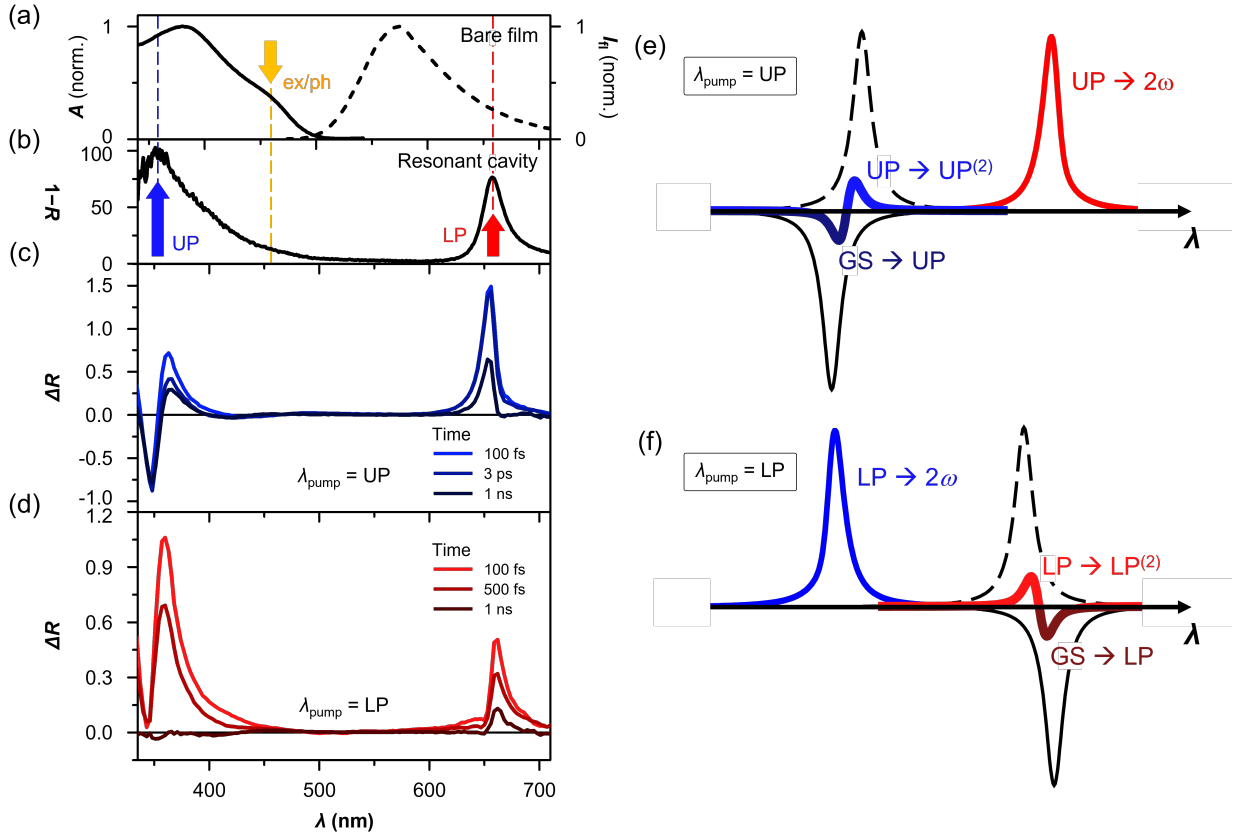

**Figure 4.** (a) Steady-state absorption and fluorescence spectra of 4CzIPN and (b) steady-state reflectance spectrum of a resulting strong-coupled cavity. The yellow arrow with a dashed line behind indicates $S_0$–$S_1$ transition energy, where the cavity photon mode is tuned to achieve strong coupling, and the blue and red arrows and corresponding dashed lines indicate the peak positions of the upper and lower polaritons. Transient reflectance spectra of strong-coupled cavity upon photoexcitation at (c) upper (UP) and (d) lower polariton (LP) at representative time delays. Since the cavity has a thick bottom mirror layer that permits no transmission, changes in the reflectivity are directly related to the changes in the absorption; $\Delta R > 0$ to excited-state absorption and $\Delta R < 0$ to ground state bleach. Tavis-Cummings model predicts transitions from one-particle to two-particles polariton states, which can be simulated into spectral signatures of (e) upper and (f) lower polariton population. The single Lorentzian-shaped excited-state absorption corresponds to a transition from UP/LP to $2\omega$ state (red and blue solid lines, respectively). The derivative-like shape is generated by adding ground state bleach (thin solid black lines) to excited-state absorption from UP/LP to $UP^{(2)}/LP^{(2)}$ (thin dashed black lines). Details of the transitions are detailed in the main text.

Under our experimental conditions, the Rabi splitting was independent of the pump intensity, meaning that we probed the vacuum Rabi splitting (limit of zero and single photons in the polariton states) where the pump excited the system from the ground to the one-particle states. Using the Tavis-Cummings model (Equation 2), we rationalized our findings as ESA processes originating from 'one-particle' LP, UP and dark states to 'two-particle states' (Figure 5). This explanation is analogous to that for exciton to biexciton transitions in semiconductors[94] and exciton to two-exciton ESA in molecular systems.[95]

At early times, ESA from polariton states dominate, but at later times dark states are eventually populated from polariton states due to energy transfer and contribute to the transient spectra as well. In particular, we identified the early time ESA at the UP (LP) region upon photoexcitation of the LP (UP) as an ESA from the LP (UP) to a state at twice the energy of the cavity mode (ESA from LP/UP → $2\omega$). After selective photoexcitation of the LP, this ESA (LP → $2\omega$) completely decayed after 1 ps, and we therefore assigned this 1 ps timescale to the lifetime of the LP.

We also identified an ESA from the UP to a two-particle state UP$^{(2)}$ at an energy slightly lower than twice UP energy, which, combined with the UP GSB gives rise to the derivative shape observed in the UP region upon excitation of the UP. A similar derivative shape was expected due to ESA from the LP to the two-particle LP$^{(2)}$ state at an energy slightly higher than twice the LP energy, however no such derivate-shape was observed, suggesting that the ESA in the LP region was dominant.

The ESA transitions probe the two-particle Rabi splitting and extend to a large number of molecules the characteristic $\Omega_R^{(2)} \sim \sqrt{2}\Omega_R$ relation between the one-particle and two-particle Rabi splitting that has been observed for a single atom.[27,58] ESAs from dark states to two-particle states DLP$^{(2)}$ and DUP$^{(2)}$ at energies slightly lower than the UP energies, and slightly higher than the LP were also identified, explaining the similarities in the spectra of the UP at short times with those at long times (i.e. ns) when the UP population is expected to have decayed into dark states.

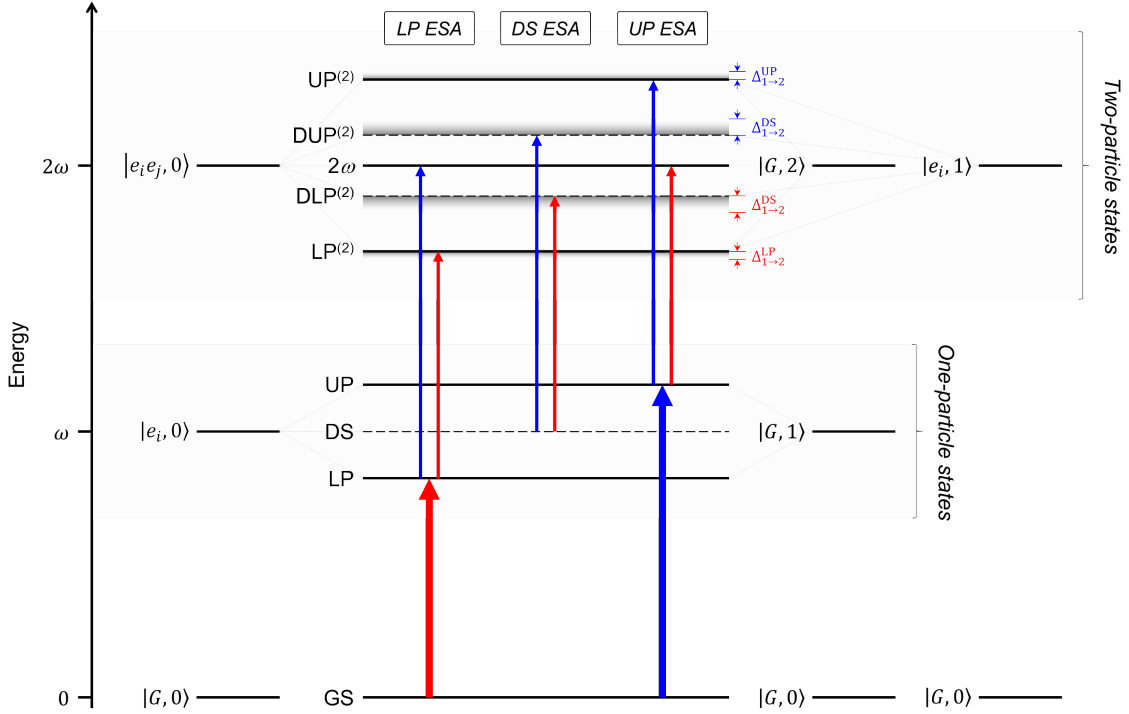

**Figure 5.** Energy structure of the one- and two-particle states in the Tavis-Cummings model for a resonant cavity of frequency $\omega$. Light-matter states with only excitons ($|e_i, 0\rangle$, $|e_i e_j, 0\rangle$), only photons ($|G, 1\rangle$, $|G, 2\rangle$), and one exciton and one photon ($|e_i, 1\rangle$) are plotted at their respective energies. Coupling of states with one exciton and no photon ($E = \omega$; left) to states containing no exciton and one photon ($E = \omega$; right) creates one-particle states, and coupling between states with two excitons and no photon ($E = 2\omega$; left), one exciton and one photon, and no exciton and two photons ($E = 2\omega$; right) give rise to two-particle states. LP (LP$^{(2)}$) and UP (UP$^{(2)}$) denote lower and upper polaritons in the one(two)-particle manifold. DLP$^{(2)}$ and DUP$^{(2)}$ are two additional states in the two-particle states, which are dark from the one-particle polariton states. Dipole allowed transitions are shown as arrows (red and blue correspond to transitions at or close to the energy of LP and UP respectively).

Interestingly, when photoexciting the LP the ESA in the UP region, also survived in the ns timescale, well beyond the estimated 1 ps LP lifetime. In addition, the evolution of this ESA at long times was very similar to the time evolution of the same ESA upon excitation of the UP and coincided with the decay of the bare molecules. A possible explanation for this observation is exciton scattering/relaxation from the LP to the dark states, which is quite surprising given the huge energy gap (0.8 eV). As we will discuss more in detail later on, the large density of dark states can explain this counterintuitive uphill energy transfer, since even though the probability of reaching any of these states by thermal excitation is very small, the high number of target states makes it possible.

In sum, our study of cavity-confined 4CzIPN evidenced the potential of pump probe experiments for obtaining the energy structure that is accessible via 1- and 2-photon absorption processes as well as the ultrafast energy transfer among polaritons and dark states. However, it also shows the challenges faced to decouple the different energy pathways, despite the seemingly simple transient spectra since many of the allowed optical transitions overlap in energy.

***Ultrafast response of vibrational polaritons.*** Owrutsky, Simpkins and coworkers recently investigated the transient response of sodium nitroprusside (SNP) with its NO vibrational band resonantly coupled to an optical cavity.[91] In this study they used both two-dimensional infrared (2DIR) spectroscopy and pump-probe spectroscopy to gain information on the energy structure and dynamics of the cavity-coupled SNP. 2DIR spectroscopy is in many ways similar to pump probe, but three rather than two pulses are used such that by varying the time-delays between the pulses, 2D maps as a function of excitation and probe frequencies at a given waiting time between second and third pulses can be obtained.[96] Cuts along a fixed pump frequency yield the standard transient absorption spectra from pump-probe, but with the advantage that correlations among the excitation and pump frequencies can be analyzed yielding information about the nature of disorder (i.e. homogeneous vs inhomogeneous) and coupling as well as energy transfer between states.

Using 2DIR and pump-probe with either broad excitation or selective excitation of the LP, UP and uncoupled reservoir molecules, the authors were able to distinguish between polariton specific responses that were present at short times (~8 ps) and reservoir response present at short and long times (>20 ps). In particular, the short-time transient spectra were markedly different depending on whether the pump excited the uncoupled molecule $\nu = 0\text{-}1$ NO vibrational transition or either polariton. In the long-time limit, all spectral shapes were almost equivalent, strongly suggesting that the shorter time response involved polariton transitions. The reservoir transient absorption spectra showed a derivative feature in the UP region, that was explained as resulting from a contraction of the Rabi splitting after a fraction of the molecules were excited to the $\nu = 0\text{-}1$ transition with the pump, plus an ESA near the LP region arising from a transition from the $\nu = 1$ to the $\nu = 2$ vibrational excited states. Kinetics showed that the reservoir population decayed in the 20 ps timescale, almost identical to the excited vibrational lifetime outside of the cavity.

In order to understand the polariton related transitions, a generalization of the quantum Rabi model for vibrational polaritons was used and the predicted transitions agreed well with those observed in the experiment. Similar to our analysis of transitions between the one- and two-particle polariton states in our electronically coupled 4CzIPN, the authors found evidence of transitions from the UP/LP to a higher excited $UP^{(2)}/LP^{(2)}$ polariton state. The UP to $UP^{(2)}$ transition spectral feature contained two decay times of 1 and 8 ps, much longer than the cavity photon lifetime and molecular lifetime, and interestingly, longer than the dephasing time observed when exciting a superposition of LP and UP with a broad pulse. The authors suggested that such a long UP lifetime, longer than the dephasing time, indicated that the UP may also decohere within its lifetime giving rise to a mixture of incoherent and coherent populations, although decoherence between different polariton states could well be faster than within each polariton.

As in our experiment with 4CzIPN, spectral congestion, here in the vicinity of the LP due to both LP and reservoir transitions, prevented a clear determination of the LP lifetime and whether its dynamics are influenced by the interplay with other excited states. Regarding reservoir molecules though, it seems unlikely that there was any energy transfer to the LP given the similarity in the dynamics when pumping the reservoir molecules with those observed outside the cavity.

***Dark states, uncoupled states, and their interplay with polaritons: modeling the free energy landscape.*** As we have discussed, the strong coupling of light and many molecules presents a unique opportunity, specifically we see the immense correlation and delocalization across distant molecules to form the polariton states. Notably, we also see the formation of the dense manifold of dark molecular states resulting from the very fact that a large number $N$ of molecules are participating. Specifically, in the one-particle subspace, the creation of the LP and UP states implies the formation of $N$-1 dark states. As noted, not every molecule in the cavity will participate in strong coupling,[35,77,78] thus $N$ here is only representative of the coupled molecules as the uncoupled population does not contribute to the formation of the polariton states or dark states. Similar to the dark states, this uncoupled population can also interact with the UP and LP.

While many theoretical studies[40,42,80,97-101] and experiments[20,43,76,87,102,103] have addressed the role of dark states and uncoupled molecules in the dynamics of strongly coupled molecules, there are still no clear guidelines to allow one to a priori understand if dynamics will be limited or even dominated by dark states or uncoupled molecules. For example, in the study on vibrational polaritons that we have previously discussed,[91] there was no evidence that the uncoupled molecules would exchange energy with the LP. However, our analysis of the ultrafast dynamics of 4CzIPN in the ultrastrong electronic coupling regime, suggested that the dark states are populated spontaneously following photoexcitation of the LP state.[20] This was surprising considering the LP state was 0.8 eV below the peak of the dark state band in this experiment.

To assess the conditions of energy transfer between dark states and polaritons, and in particular, the uphill energy transfer from the LP state to the dark states, we recently investigated the energy landscape of the one-particle states predicted by the Tavis-Cummings model (Equation 2) in terms of the *free energy*.[104] We predicted that when a very large number of molecules are involved in polariton formation, the entropic gain in the degenerate dark state manifold results in the possible reordering of the states.

In this study, the free energy of a state is given by

$$\Delta G = \Delta E_{0-j} - T\Delta S_{0-j} \qquad \text{(Eq. 8)}$$

which states that the transition from the ground state 0 to excited state $j$ is accompanied by a change in free energy that includes not only the electronic energy, $\Delta E_{0-j}$, but also the entropy term, $-T\Delta S_{0-j}$. The latter is often disregarded because entropy does not usually vary much among molecular state transitions. However, we have shown that this is no longer the case when we are considering the transitions between polariton states and the dense manifold of dark states.

The entropy of a general quantum state $j$, characterized by its density matrix ρ, can be calculated using the von Neumann entropy as defined by Equation 7. In Equation 7, the entropy is defined in units of bits. Multiplication of Equation 7 and the constants $k_B \ln 2$, where $k_B$ is the Boltzmann constant, provides entropy in units of energy per Kelvin as is standard for calculating free energy in Equation 8.[60,105] If we consider disorder in the molecular site energies (as in Figure 2) and analyze the entropy of the ensemble-averaged density matrices, we find that the entropy of the LP and UP decreases with increasing number of molecules, $N$. On the other hand, we find that the entropy of the dark states increases as $N$ increases, which also agrees with the known relationship between degeneracy and entropy. This indicates that for a given Rabi splitting, the free energy of the DS will reorder below that of the LP if N is large enough.

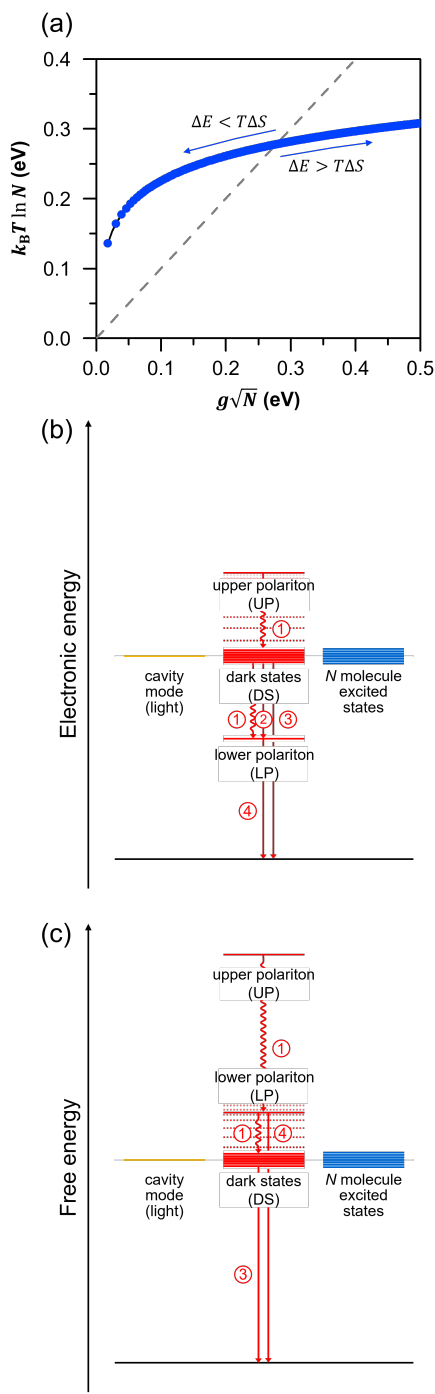

**Figure 6.** (a) The entropy against the electronic energy difference between the LP and DS for varying number of emitters (blue dots) at room temperature ($T = 300$ K) with a coupling strength of $g = 1.24$ meV ($10$ cm$^{-1}$). Gray dashed line indicates the entropic term is equal to the electronic energy difference in Eq. 8. In the regime where entropic gain surpasses the energy gain ($\Delta E < T\Delta S$; left arrow) left arrow), the DS are located below the LP state in the free energy scale, while outside the regime

($\Delta E > T\Delta S$; right arrow), the DS are located above the LP. As *N* decreases, it is important to understand that a critical minimum *N* will be reached for this fixed *g*, in which the evaluation of reordering becomes meaningless as strong coupling and polariton formation are no longer feasible. (b) Electronic energy and (c) free energy levels of cavity mode (yellow line), excited-states of molecules (blue lines), and polariton manifold (red lines) as a result of strong coupling. Nonradiative (wavy arrows) vibrational relaxation (process 1) effectively delivers population from and to DS, which act as a trap. In a conventional view in the electronic energy landscape, DS population either directly decay to the ground state (process 3) giving off emission or decay to lower polariton (LP) state via radiative (solid line) pumping (process 2) and nonradiative vibrational relaxation (process 1). LP relaxes to the ground state by emitting a

While increasing *N* results in increased entropy of the dark states, it is critical to note a competing relation. We recall that the energy difference between the dark states and the LP or UP is estimated as $g\sqrt{N}$ or half of the Rabi splitting. One can see that as *N* increases the energy difference between the dark states and the polariton states increases as well. This suggests that as *N* increases and the energy difference grows, the transfer from the LP to the dark states becomes less favorable in the traditional electronic energy ordering.

However, in accordance with Equation 8, there is a regime under which the entropic gain surpasses the gain in energy difference between the dark states and LP, and the states reorder with respect to free energy. This is illustrated in Figure 6a, where the single molecule light-matter coupling *g* is kept fixed at 1.24 meV (10 cm$^{-1}$) such that varying the number of molecules *N*, modifies both the electronic energy gap between DS-LP and the entropy of the DS. Specifically, up to a critical maximum *N*, we see the LP located above the dark states in the free energy landscape, thus inverting the usual spectroscopic energy order and allowing the LP to transfer energy (and decohere) into the dark states (Figure 6b and 6c). While this reordering can have a profound effect on the transfer between the LP and the dark states, the population dynamics following photoexcitation of the UP are not expected to be significantly altered. The UP remains higher in spectroscopic and free energy than the LP and dark states, thus it is still favorable to transfer energy from the UP to these states. Overall, this view based on free energy considerations is consistent with the picture of population equilibration under thermal equilibrium.

Under thermal equilibrium, one assumes the transitions in the cavity follow ergodic Markov Chain dynamics. Subsequently, one may use detailed balance to compare the energy transfer between the LP and dark states such that

$$k_{LP \to DS} = k_{DS \to LP}(N-1) \exp\left((E_{LP} - E_{DS})/k_B T\right) \qquad \text{(Eq. 9)}$$

where $k_{i \to f}$ is the rate constant describing the transfer from the initial to final state. Here, note that the degeneracy of the dark states is accounted for, and a Boltzmann probability distribution is used. Equation 9 here effectively results in the comparison of the spectroscopic energy gap $\Delta E$ and $k_B T \ln(N)$, the latter being the classical entropic contribution to the free energy (Equation 8) for a state with degeneracy $N$.

Using this relation, our group further evaluated the dynamics between the dark states and the UP and LP using the model described in Figure 7a. In this model, the ratio between the rate constant for the transfer from the DS to LP and the rate constants for transitions from the UP was 1:100.[103] This ratio was primarily used to describe the significantly faster nature of transfers from the UP.[22,103] Using a generalized form of Equation 9, the rate constants of the respective uphill transitions could then also be determined, where the energy of the UP (LP) states was defined as $\pm g\sqrt{N}$ and the temperature was 300 K. In this model, the molecule-light coupling parameter, g was held constant at 1.24 meV (10 cm$^{-1}$) and the primary focus was on the dynamics and stationary distributions following the photoexcitation of the LP.

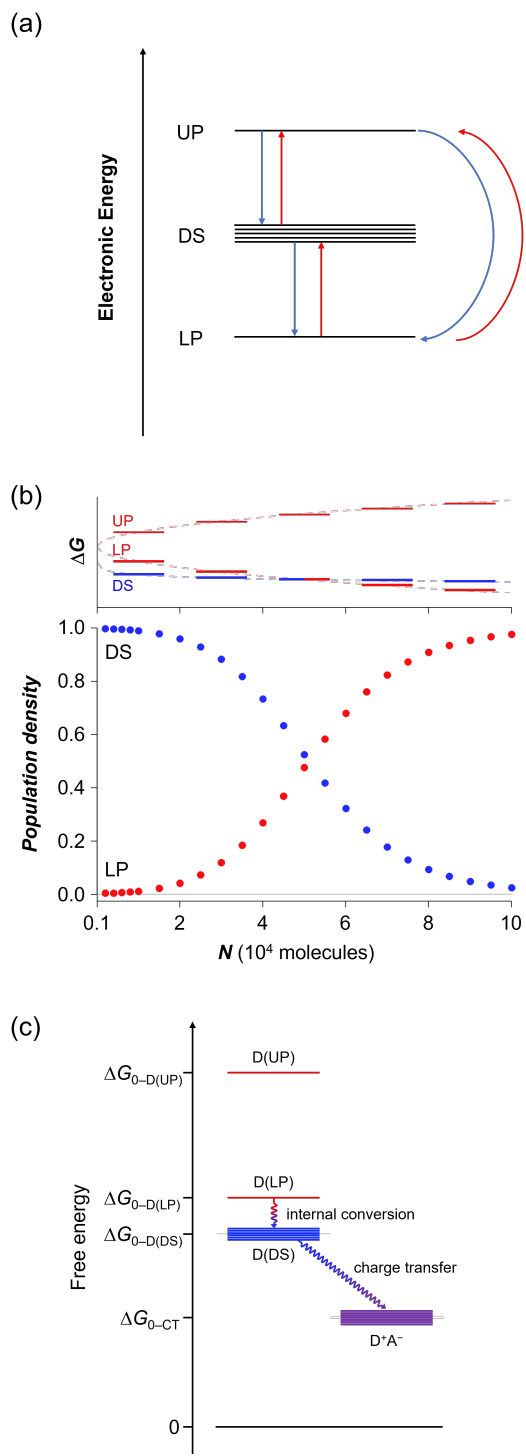

**Figure 7.** (a) Electronic energy ordering of the lower polariton (LP), dark states (DS), and upper polariton (UP). Blue arrows signify the known favorable downhill transitions with given rate constants. Red arrows describe the uphill transitions believed to be unfavorable with respect to electronic energy ordering. Rate constants for transitions

described by red arrows are determined by Eq. 9. (b) The free energy ordering of UP, LP (red bars), and DS (blue bars) (upper panel) and stationary distributions of the DS (blue dots) and LP (red dots) populations following photoexcitation of the LP (lower panel) for $N$ ranging from 2,000 to 100,000 molecules with $g =$ 1.24 meV (10 cm$^{-1}$) at a temperature of 300 K. Note the UP was not populated following photoexcitation of the LP as the dynamics from the UP are significantly faster, thus making the uphill transition even more unfavorable. (c) The free energy diagram of polariton states of a donor molecule in a donor-acceptor charge-transfer complex. When the LP is higher in free energy than the DS, the excited-state population generated by $\Delta E_{0-D(LP)}/\hbar$ can be spontaneously transferred to DS, then to the low-lying charge transfer state.

The stationary distributions for varying numbers of molecules, specifically $N$ ranging from 2,000 to 100,000 generic emitters, can be seen in Figure 7b. One can see the competition between the entropic gain (approximately logarithmic in $N$) and the energy-lowering of the LP state (proportional to $\sqrt{N}$) as $N$ increases. Specifically, as $N$ increases, a LP–DS energy difference is reached where the LP becomes more favored than the dark states and the uphill transfer becomes less probable (Figures 6a and 7b). In other words, after a critical maximum number of molecules has been reached, adding more molecules to the cavity no longer results in a reordering of the DS below the LP in terms of free energy.

Prior to this critical energy difference and for the $N$ and $g$ values considered, the entropic gain of the degenerate dark states prevails, and the dark states are lower in free energy than the LP state (Figure 7b). It is paramount to realize that due to $g$ being fixed there also exists a critical minimum number of molecules, in which, below this point, strong coupling and polariton formation are no longer possible and the evaluation of reordering becomes unrealistic. It is then within this allowed reordering regime that the transfer from the LP to the dark states is favored even though this transition is traditionally uphill in terms of electronic energy. Overall, this possibility of achieving a quasi-equilibrium between the LP and dark states provides a clear working mechanism for uphill population transfer from the LP and is able to explain spectroscopic observables, such as temperature-dependent anti-Stokes fluorescence.[37]

While this model treated the dark states as degenerate, disorder or variability in the system can result in this degeneracy being broken and the energies of the dark state manifold broadening.[34,39,77,100,101,106] For small system disorder, such as small variations in molecular transition frequencies, molecule positions, and dipole orientations, we expect the degenerate model used here to still accurately describe the transitions of interest within the cavity. This is because the energy gap between the LP state and the average of the dark state energies is larger than the broadening caused by disorder. Furthermore, the increased probability of accessing the lowest energy of the dark states from the LP is counterbalanced or averaged out by the decreased probability of accessing the highest energy of the dark states from the LP. Thus, for small disorder, we propose similar dynamics and results as those observed above. For cases where there is significant overlap between the LP and DS peaks, i.e. large disorder or small Rabi splitting, Groenhof and coworkers found that the degree of overlap is indicative of the accessibility of the dark states, with increased overlap pointing towards a larger dark state population following excitation of the LP.[100]

As mentioned, uncoupled molecules or states, such as the triplet state, also have the potential to interact with polariton states. Therefore, the proposed reordering of the dark state free energy below that of the LP also has consequences on how excitation will flow to and from these relevant excited states that are not coupled to the cavity mode. Chemical reactivity is often boosted by photoexcitation because the energy of an absorbed photon significantly changes the driving force for transforming reactants to products.[107-109] When quantifying this change in driving force, it has generally been assumed that the entropy change accompanying photoexcitation of molecular systems is small. Models, such as theories for photoinduced electron transfer, therefore equate the free energy changes from reactants to products with energy changes that can be measured directly using spectroscopy.[110] Generally, therefore, our intuitive view is that population dynamics flow spontaneously "downhill", where the ordering of states is prescribed by the ladder of electronic energy gaps. However, this intuitive connection between the spectroscopic ordering of states and the spontaneity of reactivity breaks down dramatically as shown for strongly coupled polariton states. Thus, the exceptionally low von Neumann entropy of coherent states arising from strong coupling presents an unforeseen resource.

Where and how might we exploit the entropy-differential resource? One application is to use the entropy difference to supplement photoexcitation energy, thereby achieving photochemical or photophysical transformations with much lower excitation energy than in the absence of strong light-matter coupling. The concept is the same as that exploited in any spontaneous endothermic transformation. For example, chemists are familiar with solvation of ammonium chloride in water, that occurs readily, despite unfavorably $\Delta H$ for solvation, by a compensating $\Delta S$. The entropic drive is evidenced by cooling of the solution.

To give a concrete example, consider the Rehm-Weller equation.[110] This relation is commonly used to predict the energetic driving force for photoinduced electron transfer from a donor D to acceptor A that are separated by a distance $a$ in a solvent with relative dielectric constant $\varepsilon$ and at temperature $T$:

$$\Delta G = \text{IP}_\text{D} - \text{EA}_\text{A} - \frac{e^2}{(4\pi\varepsilon_0)\varepsilon a} - \left(\Delta E_{0-\text{D(LP)}} - T\Delta S'\right), \qquad \text{(Eq. 10)}$$

Here, IP$_D$ is the ionization potential of the donor, EA$_A$ is the electron affinity of the acceptor, $\varepsilon_0$ is the vacuum permittivity, and $\Delta E_{0-D(LP)}$ is the transition energy from the ground state to the donor LP state. $\Delta S'$ is the difference in entropy between the donor LP and any fully mixed state, such as the ground state to the LP (which is explicit in the original Rehm-Weller equation) or the LP to dark states (which is approximately the same as that between the LP and ground state). In the original paper, Rehm and Weller note that for typical bimolecular charge transfer reactions, the entropy term will be negligible, and can be safely ignored. We hypothesize that this is not the case for polariton-modified systems.

If we assume relaxation from the LP to dark states is fast compared to the rate of electron transfer, and that we have conditions such that the system equilibrates to a majority excited state population of the dark states, then we can make the approximation that $\Delta E_{0-D(LP)} - T\Delta S' \sim \Delta E_{0-D}$, that is, the donor excitation energy outside the cavity. Therefore, despite exciting the system with light of frequency $\Delta E_{0-D(LP)}/\hbar$, which can be ~2000 cm$^{-1}$ (~250 meV) below the excitation energy of the same molecules outside the cavity, the effective excitation energy added to the system is $\Delta E_{0-D}$. Therefore, the Rehm-Weller equation (Equation 10), with account of the entropy change, predicts that the driving force for photoinduced electron transfer in a cavity is the same as if the molecules were outside the cavity. This idea can be exploited in solar photochemistry and photoredox chemistry, where we imagine that strong light-matter coupling can be used to enable, for instance, blue excitation wavelengths to leverage high redox potentials of catalysts that normally absorb in the near ultraviolet. Particularly in photoredox chemistry,[107,111-114] this is a strategy that might make more difficult bond activation chemistry within reach.

Another application where the entropic component of dark states free energy may be exploited is to alter the rate and efficiency of the reverse intersystem crossing (RISC) process of thermally activated delayed fluorescence (TADF) in molecules. The idea is that the energy of the newly formed polariton states can be tuned with respect to that of the uncoupled triplet state, possibly even inverting the triplet and singlet energies (Figure 1d). A recent experiment by Börjesson and co-workers, indeed demonstrated the concept by showing modification of RISC rates by varying the energy of polariton states with respect to triplet states and dark states.[46] In one case where the lower polariton was lower in energy to the triplet state ($E_{LP} - E_{T1} = -68$ meV), authors found that the rate of delayed emission became independent of temperature, an indication of a direct barrierless energy transfer. This is in contrast to other works that have found that energy flows predominantly between the triplet and the purely molecular dark states rather than the LP, which leaves RISC rates unaltered by strong coupling.[26] We suggest that an analysis in terms of free energy as the one we have proposed above could provide further insight into how barrierless transitions are attained despite obvious energy gap and, more generally, when the LP is expected to participate in RISC to modify its rate and efficiency.

***Conclusions.*** In this perspective, we discussed some of the present theoretical and experimental efforts occurring in the dynamic field of polariton chemistry with a focus on the collective aspects of polariton formation in molecular systems and within this context, the largely open question of how energy flows in such systems. In particular, we have illustrated how strong coupling emerges as a consequence of the coupling of many molecules to a common optical mode, which relaxes the need of strong coupling of single molecules. This however comes at the expense of the formation of a large density of molecular dark states which in principle interact with polaritons as do other purely uncoupled molecular states. Understanding these interactions is key to control molecular function via strong coupling.

In this respect, we have illustrated the capabilities and benefits of spectroscopic techniques, such as ultrafast pump-probe spectroscopy. One such benefit is the ability to assess the interaction between polariton states and uncoupled molecular states by resolving respective spectral signatures in the frequency and time domain. Multidimensional spectroscopy also presents itself as a promising technique in this regard as its double-pulse pump sequence, which grants a pump-frequency resolution, enables visualization of interstate correlations via off-diagonal elements in the two-dimensional spectra.[96] A couple of recent reports have shown that this technique is well-suited in detangling the key dynamics in electronic strong coupling systems, such as relaxation pathways from polariton to dark states[103] and system-bath interaction in polariton states.[62]

Despite these stimulating results, more studies that examine the coupling between polariton and, for example, triplet states, charge transfer states, and photoproduct states are needed. The knowledge from these experiments will provide not only a better understanding of the polaritonic system, but also a path towards other kinds of more precise control of polariton-modified dynamics in the future. Specifically, use of quantum control schemes either by including additional pump pulses as in a pump-repump sequence or by employing shaped pulses[115-117] in ultrafast pump-probe spectroscopy could reveal how cavity dynamics, and ultimately desired reactions, might be manipulated.

Throughout this perspective, we also discussed the role of entropy and free energy in polariton dynamics. We now have increased understanding of the role of entropy in the cavity and the competition between electronic energy and entropy in the reordering of states with respect to free energy, which can determine the extent to which dark states are populated as thermal equilibrium is reached. While equilibration to dark states seems to imply that strong coupling does not modify molecular processes (e.g. their rates), it may still be used as a resource if the LP lies above the dark states in free energy. This would allow, for example, the execution of reactions with energy less than that required for the molecules without strong light-matter coupling. Further experiments designed to check the proposed reordering of dark states and lower polariton by varying the number of molecules that couple to the cavity mode or tuning temperature would be timely and would provide new guides to exploit strong light-molecule coupling.

**ACKNOWLEDGMENTS**

This work was funded by the Gordon and Betty Moore Foundation through Grant GBMF7114. F.F. acknowledges financial support from the European Union's H2020 Marie Skłodowska Curie actions, Grant agreement No 799408.